\documentclass{article}

\usepackage{arxiv}

\usepackage[utf8]{inputenc} 
\usepackage[T1]{fontenc}    
\usepackage{hyperref}       
\usepackage{url}            
\usepackage{booktabs}       
\usepackage{amsfonts}       
\usepackage{nicefrac}       
\usepackage{microtype}      
\usepackage{graphicx}
\usepackage{natbib}
\usepackage{doi}
\usepackage{amsmath}
\usepackage{siunitx} 
\usepackage{makecell} 
\usepackage[labelformat=empty]{subfig}
\usepackage{dirtree}
\usepackage{multirow}

\title{Consumer-grade EEG-based Eye Tracking}

\author{\href{https://orcid.org/0009-0006-3090-2955}{\includegraphics[scale=0.06]{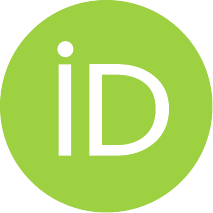}\hspace{1mm}Tiago Vasconcelos Afonso} \\
	Department of Mathematics and Natural Sciences \\
	Darmstadt University of Applied Sciences \\
	Darmstadt, 64295, Germany \\
	\texttt{tiago-afonso@t-online.de} \\
	\And
	\href{https://orcid.org/0009-0003-9134-2646}{\includegraphics[scale=0.06]{orcid.pdf}\hspace{1mm}Florian Heinrichs} \\
	Department of Medical Engineering and Technomathematics \\
	FH Aachen - University of Applied Sciences \\
	Jülich, 52428, Germany \\
	\texttt{f.heinrichs@fh-aachen.de} \\
}

\renewcommand{\shorttitle}{Consumer-grade EEG-based Eye Tracking}

\hypersetup{
pdftitle={Consumer-grade EEG-based Eye Tracking},
pdfsubject={cs.LG},
pdfauthor={Tiago Vasconcelos Afonso, Florian Heinrichs},
pdfkeywords={EEG data, Eye-tracking, Consumer-Grade EEG},
}

\begin{document}
\maketitle

\begin{abstract}
    Electroencephalography-based eye tracking (EEG-ET) leverages eye movement artifacts in EEG signals as an alternative to camera-based tracking. While EEG-ET offers advantages such as robustness in low-light conditions and better integration with brain-computer interfaces, its development lags behind traditional methods, particularly in consumer-grade settings. To support research in this area, we present a dataset comprising simultaneous EEG and eye-tracking recordings from 113 participants across 116 sessions amounting to 11 hours and 45 minutes of recordings. Data was collected using a consumer-grade EEG headset and webcam-based eye tracking, capturing eye movements under four experimental paradigms with varying complexity. The dataset enables the evaluation of EEG-ET methods across different gaze conditions and serves as a benchmark for assessing feasibility with affordable hardware. Data preprocessing includes handling of missing values and filtering to enhance usability. In addition to the dataset, code for data preprocessing and analysis is available to support reproducibility and further research.
\end{abstract}


\section{Background and Summary} 

EEG-based eye tracking (ET) is emerging as a promising application of brain-computer interfaces (BCIs) \citep{dietrich2017, fuhl2023, kastrati2021, sun2023}. While EEG is typically used to record the electrical activity of the brain, it also captures eye movement artifacts due to the inherent electrical charge of the eyes. Although these signals are usually considered noise in other BCI applications and are often removed \citep{croft2000}, they can be effectively used to track eye movements. These signals are also easier to decode than brain activity, as they are not complicated by the complexity and noise associated with brain signal interpretation. In addition, achieving reliable and accurate eye tracking using EEG technology could significantly enhance existing consumer BCIs, opening up a wide range of new applications. Apart from the potential for BCI applications, EEG-based eye tracking is an interesting alternative to eye tracking in its own right, offering several advantages over camera-based eye tracking, which is the predominant method used for eye tracking today. Compared to camera-based methods, EEG-based eye tracking works even in poor lighting conditions or with closed eyes, such as during sleep. Additionally, performing experiments that require both eye movement and EEG data are more practical with EEG based eye tracking, as no additional hardware is required. Such experiments are quite common in cognitive neuroscience or psychology research. The same holds for consumer BCIs, where the integration of eye tracking is more practical with EEG-based eye tracking than with camera-based eye tracking. However, the development of EEG-based eye-tracking is still lagging behind. Previous results have been achieved under laboratory conditions with expensive hardware \citep{dimigen2011, nikolaev2016, zhang2025}. These systems are impractical for everyday use (e.g., because a gel must be applied between the electrodes and the scalp before use). 

All of this motivates the creation of a new dataset, that contains EEG and eye tracking data collected simultaneously. This dataset is intended to be comparable in structure to the already existing EEGEyeNet dataset \citep{kastrati2021}, but  based on consumer hardware with dry electrodes. Since no such dataset currently exists and the feasibility of reconstructing gaze position from EEG data recorded with consumer hardware is uncertain, the dataset is divided into different levels of complexity. The first level includes data, where eye movements occur along a single axis, either up, down, left, or right. The second level includes data with eye movements along both axes. 

Using this dataset, the performance of existing EEG-ET methods as well as new ones can be evaluated on consumer-grade EEG. This allows a more realistic evaluation of EEG-ET methods in real applications than data sets recorded under artificial conditions with high-end measurement technology in laboratories. Furthermore, this data set is intended to serve as a new, challenging benchmark for future research in the field of EEG-ET.

\section{Methods} 

The dataset consists of simultaneously recorded EEG and eye tracking data from $113$ participants, collected during $116$ sessions, in which $4$ experimental paradigms were presented to the participants. 
The experimental paradigms are designed in a way that it becomes increasingly challenging to reconstruct the gaze position from the EEG data as the eye movement becomes less restricted. In every paradigm, the participants were instructed to follow a target moving on the screen as accurately as possible with their eyes. The presentation was also designed in a way that helped the participants follow the target closely.

\subsection{Participants}
\begin{figure}
    \centering
    \includegraphics[width=0.65\textwidth]{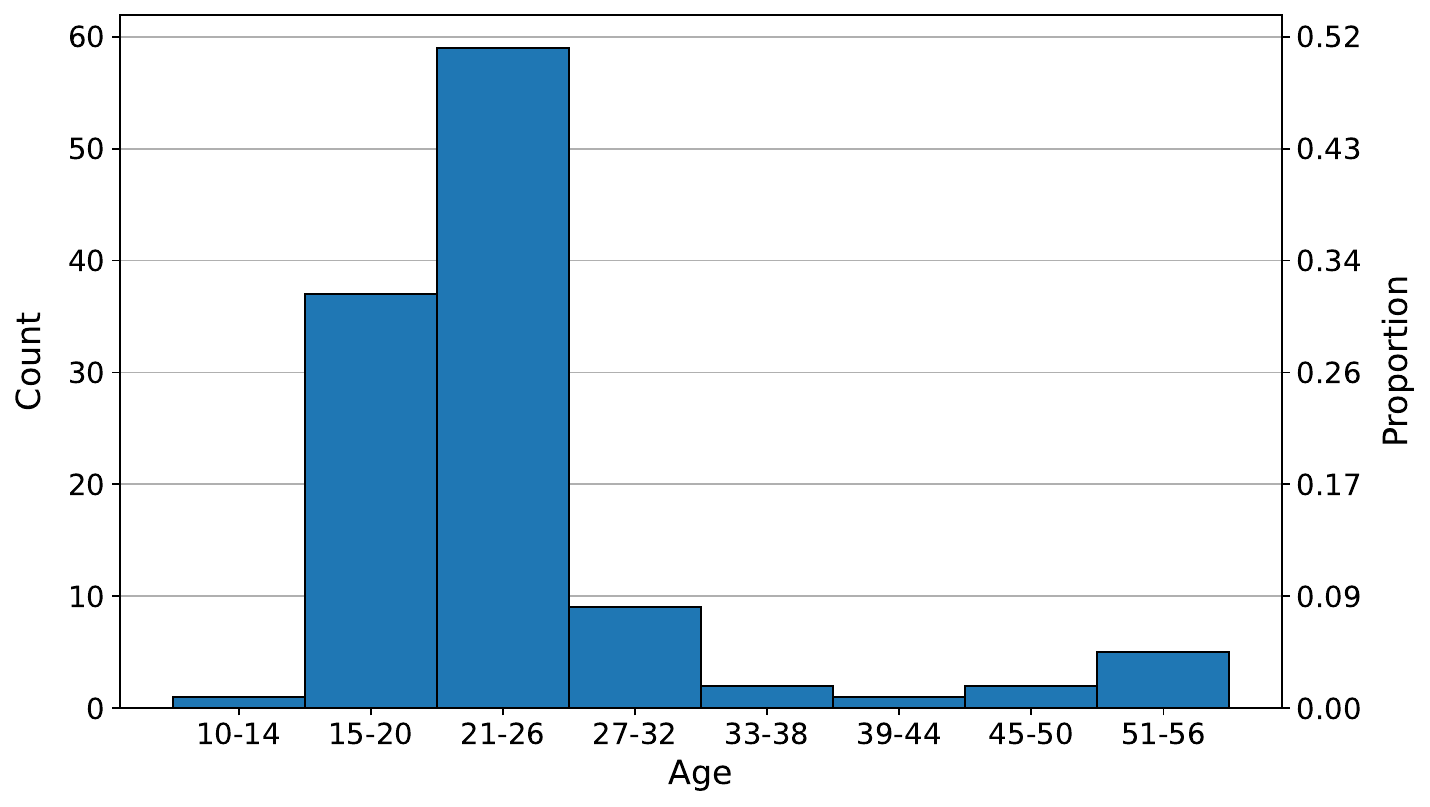}
    \caption{Histogram of the age distribution of the participants.
    The left y-axis shows the number of participants in each age group, the right y-axis shows the proportion of participants in each age group.}
    \label{fig:dataset:age}
\end{figure}

A total of 113 participants (91 males and 22 females, 100 right-handed and 12 left-handed, 1 ambidextrous) took part in this study. The group of participants is not balanced, as there were more male individuals and more right-handers. Demographic data, including the age, gender, handedness, vision correction, neurological disorders, and color blindness of the participants, is summarized in Table~\ref{tab:dataset:demographics}, the age distribution is shown in Figure~\ref{fig:dataset:age}. The participants were informed about the purpose of the study, the data that would be collected, the duration of the study, and that any collected data would be anonymized. All participants took part in this study voluntarily and received no compensation.

According to the guidelines by the Deutsche Forschungsgemeinschaft (DFG, German Research Foundation), no ethics approval was required for this study as it involved electrophysiological recordings (EEG), which did not meet the criteria that would require approval. Participants provided informed consent for participation and data sharing, and the data were collected anonymously.

Every participant was asked to sign two documents. The first document was a consent form, where the participants agreed to take part in the study, and the data collection. Additionally, participants confirmed that they were taking part in the study voluntarily and were not receiving any compensation.
The second document was the demographic questionnaire, where the participants provided information about their age, gender, handedness, vision correction, neurological disorders, and color blindness.

\begin{table}
    \centering
    \makebox[\textwidth][c]{
        \begin{tabular}{l|l|l}
            \toprule
            Demographic                          & No.\ of Participant & Percentage       \\
            \midrule
            Gender (male/female/diverse)         & $91$~/~$22$~/~$0$      & $81$\%~/~$19$\%~/~$0$\%     \\
            Handedness (left/right/ambidextrous) & $12$~/~$100$~/~$1$     & $11$\%~/~$88$\%~/~$1$\%     \\
            Vision Correction (yes/no)           & $54$~/~$59$          & $48$\%~/~$52$\%           \\
            Neurological Disorder (yes/no)       & $6$~/~$107$          & $5$\%~/~$95$\%            \\
            Color Blind (yes/no)                 & $2$~/~$111$          & $2$\%~/~$98$\%            \\
            \bottomrule
        \end{tabular}
    }
    \caption{Demographic data of the participants.}
    \label{tab:dataset:demographics}
\end{table}

\subsection{Setup}
The participants were seated in front of a \SI{24}{inch} desktop monitor with a resolution of \SI{2560}{px}$\times$\SI{1440}{px} and a \SI{60}{Hz} refresh rate, placed \SI{60}{cm} away from the participants and raised high enough such that the center of the screen matched the eye height of the subject. A webcam was mounted on a stand in front of the screen, adjusted to be as high as the lower edge of the screen, looking up at the subject. 
The recording setup is illustrated on the right of Figure~\ref{fig:dataset:setup}.

\subsection{Hardware}

\begin{figure}
    \centering

    \raisebox{-0.5\height}{\includegraphics[width=0.47\textwidth]{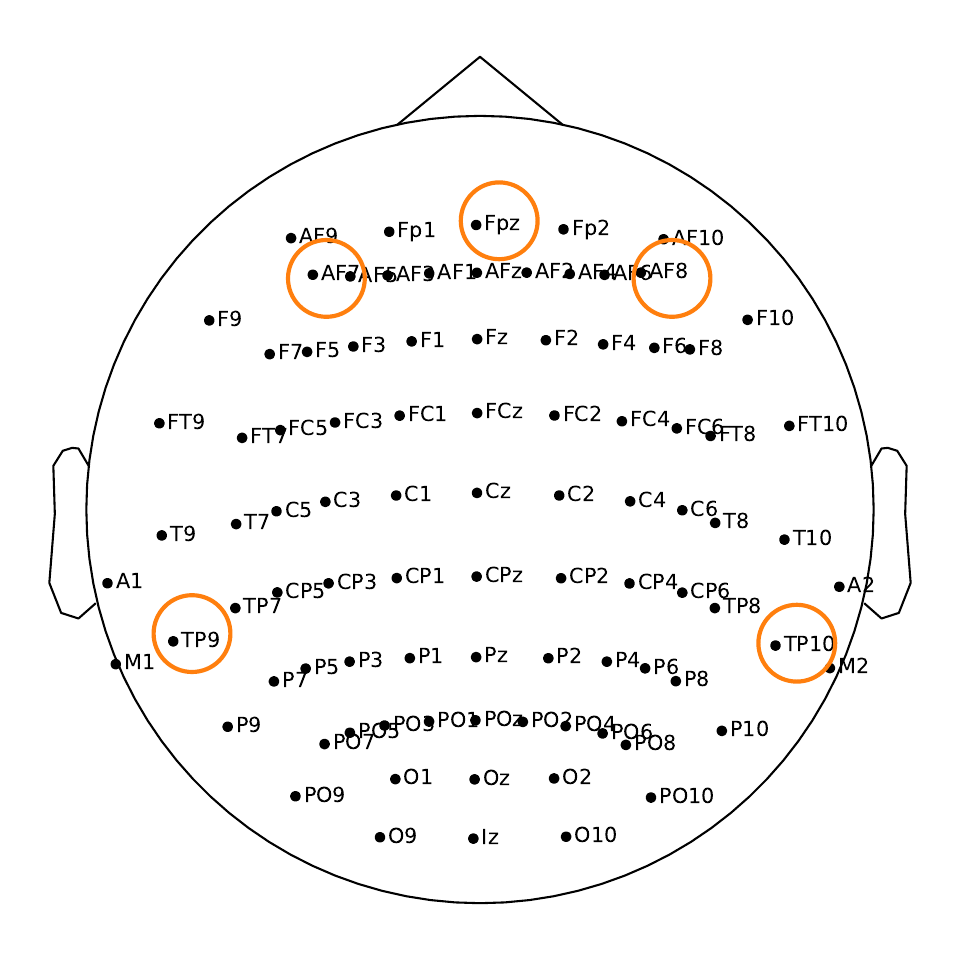}}
    \hspace{0.03\textwidth}
    \raisebox{-0.5\height}{\includegraphics[width=0.47\textwidth]{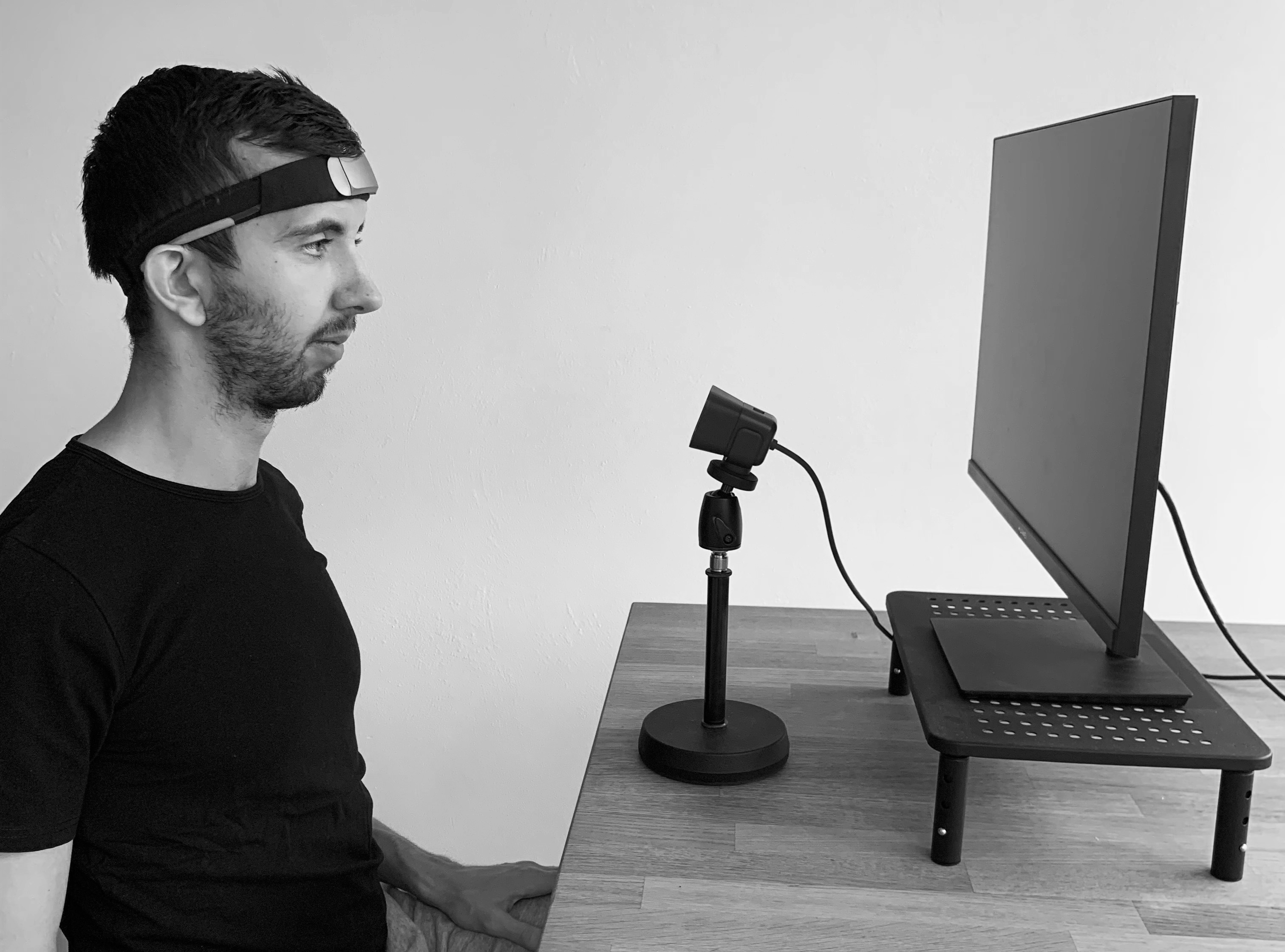}}
    \caption[Electrode Placement on the Muse S 2 Headband]{Left: Positions of electrodes of the Muse S 2 Headband in the 10-20 system. Right: Recording setup of the experiments.}
    \label{fig:dataset:setup}
\end{figure}

The webcam used was the Logitech StreamCam, which records in Full HD (\SI{1920}{px}$\times$\SI{1080}{px}), at $60$ frames per second. The monitor used was a DELL P2416D  with a 2560x1440 resolution, 59.95Hz refresh rate, and 8-bit RGB Standard Dynamic Range.
The EEG data was recorded using the Muse S 2 Headband (RRID: SCR\_014418), a consumer-grade EEG-Headset with five dry electrodes TP9, TP10, AF7,  AF8 and Fpz, where the latter is used as reference electrode. 
The headband records at a sampling rate of \SI{256}{Hz}. 
The electrode placement in the standard $10$-$20$ system is shown on the left of Figure~\ref{fig:dataset:setup}.

Two laptops were used: one for controlling the stimuli presentation and recording the data and one for showing the stimuli to the participants. Both laptops communicated via a socket connection.

\subsection{Software}

Data from three different sources was recorded: the position of the target, the gaze position, and the EEG data. The resulting three data streams were synchronized using the Lab Streaming Layer Protocol (LSL). The LSL streams were recorded using the LabRecorder software, which properly resolves the synchronization information of multiple streams. The gaze position was estimated from the webcam video using the GazePointer software.

\subsection{Experiment Procedure}
\label{sec:dataset:session}
Most sessions took place during 90 minute practice groups of different artificial intelligence related courses at the Darmstadt University of Applied Sciences. The participants were seated in the back of the room in which the practice group took place.
At the start of the group exercise, before any session started, all participants were briefed about the study and the data collection. The participants were informed about the purpose of the study, the data that would be collected, the duration of the study, that participation was voluntary, and that any collected data would be anonymized. After the briefing, the first subject was seated in front of the screen and the first session started.

Each session (nominally) lasted ten minutes in total and consisted of the following parts: introduction, practice, and recording. The introduction included the calibration of the camera-based eye tracker, fitting the EEG headset to the participant's head, and checking signal quality. Furthermore, the distance to the screen and the height of the screen were adjusted to match the eye height of the subject. The participants were instructed not to move or speak during the recording. Before each recording, a short practice session was conducted to prepare the participants for the respective experimental paradigm.

During the actual data collection, EEG and eye tracking data were recorded in four different situations: level-$1$-smooth, level-$1$-saccades, level-$2$-smooth, and level-$2$-saccades. The duration of each paradigm is shown in Table~\ref{tab:dataset:duration}. A level-$1$ experiment only required the subject to move their eyes left, right, up or down for one minute. Data from those experiments is meant to serve as a first stepping stone for potential EEG-ET techniques. 

\begin{table}
    \centering
    \begin{tabular}{l|l|l}
        \toprule
        Experimental Paradigm   & Duration & Total Duration \\
        \midrule
        level-1-smooth   & \SI{56}{\second}                 & \SI{1}{\hour} \SI{49}{\minute}    \\
        level-1-saccades & \SI{1}{\minute}                  & \SI{1}{\hour} \SI{56}{\minute}    \\
        level-2-smooth   & \SI{2}{\minute} \SI{8}{\second}  & \SI{4}{\hour} \SI{8}{\minute}     \\
        level-2-saccades & \SI{2}{\minute}                  & \SI{3}{\hour} \SI{52}{\minute}    \\
        \hline
        Total            & \SI{6}{\minute} \SI{4}{\second}  & \SI{11}{\hour} \SI{45}{\minute}   \\
        \bottomrule
    \end{tabular}
    \caption{Duration of the experimental paradigms.}
    \label{tab:dataset:duration}
\end{table}

In a ``smooth'' experiment, the target moved across the screen in a continuous manner. The subject followed the target using eye movement called ``smooth pursuit'', in which the eyes move smoothly to follow a moving object.  A ``saccades'' experiment required the subject to follow the target using a series of quick eye movements called ``saccades''. Here, the target jumped between predefined points on the screen. For all paradigms, the stimuli presentation was designed in such a way, that the user could always anticipate the movement or a jump of the target. This way, the position of the target is always close to the ground truth gaze position. If the participant experienced eye strain or fatigue, a break was taken before the next experiment.

\subsection{Stimulus Presentation}
\label{sec:dataset:presentation}

The target was presented to the participants using a custom stimuli presentation software (``Presentation App'') written in Python using the Qt framework. The bounding box in all tasks was 440\,mm x 220\,mm and the target was presented as a circle with a diameter of 10\,mm. 

\begin{figure}
    \centering
    \includegraphics[width=0.48\textwidth]{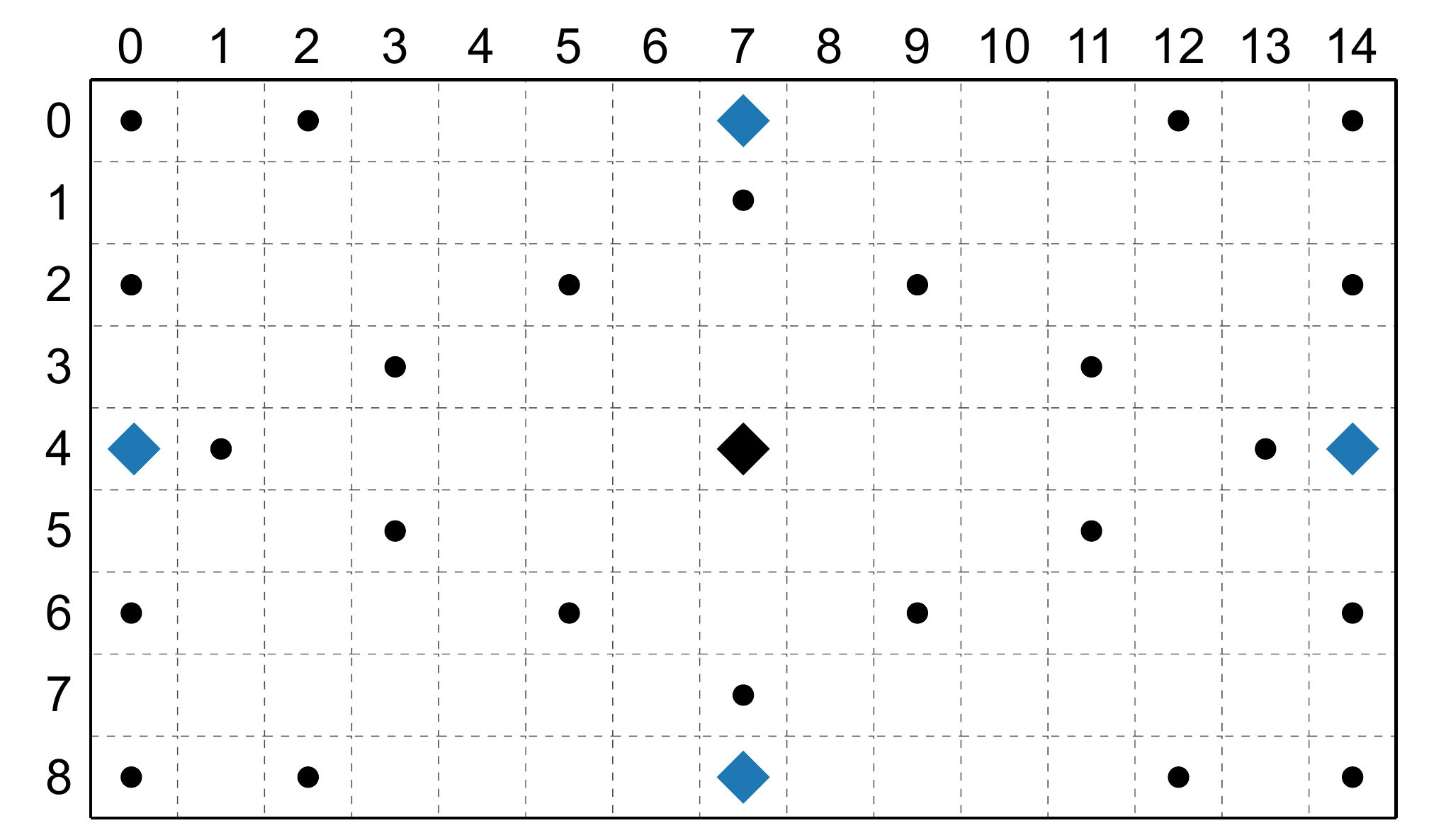}
    \includegraphics[width=0.48\textwidth]{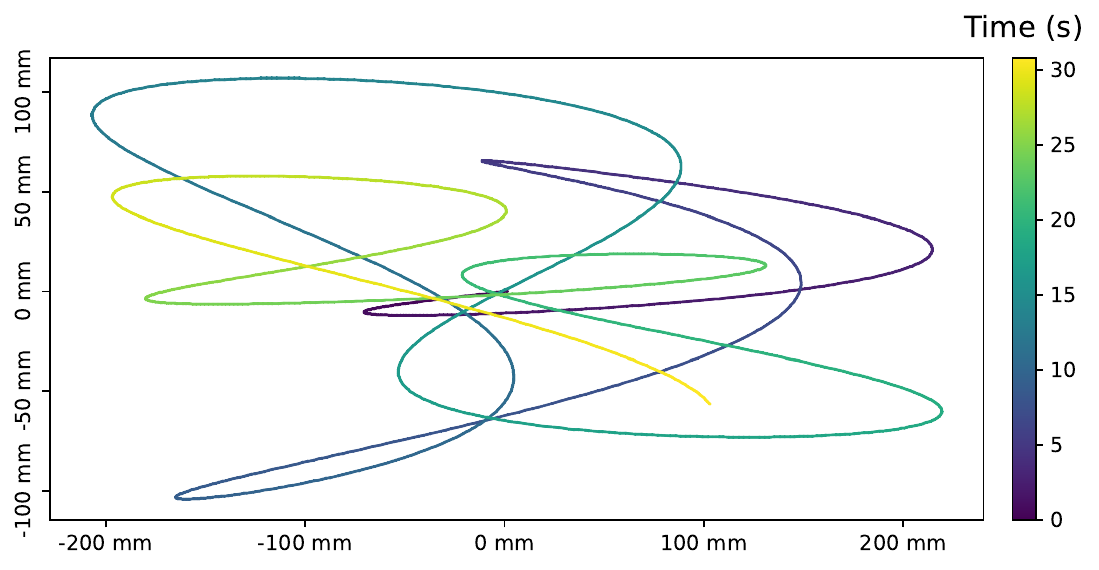}
    \caption{Left: Grid points in the ``saccades'' experiments. Displayed as blue diamonds (level-$1$ experiment), black dots (level-$2$ experiment) and black diamond (both). Right: Exemplary trajectory of the target that was presented to the participants. It was created using the parameterized function \eqref{eq:dataset:curve}.}
    \label{fig:dataset:stimulus}
\end{figure}

\textbf{Saccade Experiments.} The grid point positions in the level-$1$ and level-$2$ ``saccades'' experiments are shown in Figure~\ref{fig:dataset:stimulus} (left). In a level-$1$ experiment, there was a saccade to the edge of the bounding box every second, followed by a saccade back to the center of the screen in the next second. In a level-$2$ experiment, the target jumped to any of the grid points every $1.5$ seconds. 

\textbf{Smooth Experiments.} The path along which the target moved in the ``smooth'' experiments was created using a parameterized curve. More specifically, for a given $t \in [0,1]$, the $x$ and $y$ position of the target in millimeters relative to the center of the screen was given by:
\begin{equation}\label{eq:dataset:curve}
    f: [0,1] \to \mathbb{R}^2,\quad t \mapsto 
    \begin{pmatrix}
        \left(\cos(a t)\sin(b t) + e t\right)\cdot \frac{w}{2} \\
        \left(\cos(c t)\sin(d t) + f t\right)\cdot \frac{h}{2}
    \end{pmatrix},
\end{equation}
where $w$ and $h$ are the width and height of the bounding box in millimeters, i.\,e., $w=440$\,mm and $h=220$\,mm. In particular, the $x$-coordinate describes the horizontal and the $y$-coordinate the vertical position of the target. The time in seconds the target takes to traverse the complete curve can be controlled by the parameter $T$. The value of $t$ is incremented $120$ times per second, in steps of size $\frac{1}{120} / T$, resulting in a smooth movement. 

For all level-$1$ experiments $a = b = c = d = 0$, $e, f \in \{-1, 0, 1\}$. This results in paths where the target moves continuously from the center to one of the edges of the specified bounding box and back. For these experiments, $T = 1$, for vertical movements and $T=2$ for horizontal movements. This choice resulted in the target taking 1 second for vertical and 2 seconds for horizontal movements. The different times were chosen to account for the different height and width of the bounding box.

For level-$2$ experiments $a, b, c, d$ were randomly selected from the interval $[-50, 50]$, $e = f = 0$ and $T=28.5$, so that the target takes 28.5 seconds to traverse an entire curve. The random selection of parameters results in more complex paths on which the target moves. An example of a path created using this method is displayed in Figure~\ref{fig:dataset:stimulus} (right).
Multiple curves were shown in each experiment. Before the target started to move along the next curve, it waited for 2 seconds in the center of the screen. The value of $T$ was selected so that completing four curves and the three intervening pauses took a total of 2 minutes. 

In a level-$1$ smooth experiment, only four possible curves existed, one for each direction. The order in which the curves were shown was determined by randomly sampling without replacement from a pool where every direction existed $4$ times, resulting in a sequence of $16$ curves in total. In level-$2$ smooth experiments, four curves were shown which were created by randomly picking values for $a, b, c, d$ from the interval $[-50, 50]$. In total, $18$ different of these curves were created.

In level-$1$ saccades experiments, the target jumped $30$ times to one of the four directions (left, right, up, down) and back to the center, resulting in $60$ saccades in total. To choose the $30$ directions for the jumps, $8$ random permutations of the four directions were created, and the last two directions from the final permutation were discarded. For the level-$2$ saccades experiments, $80$ positions were selected by creating $4$ random permutations of the $25$ possible positions. The last 20 positions from the final sequence were then discarded.

\begin{figure}
    \centering
    Smooth Movement
    \subfloat[Countdown at the start]{%
        \includegraphics[width=0.24\textwidth]{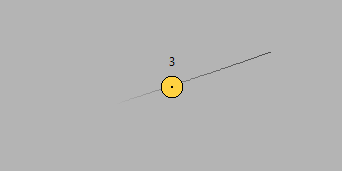}
    }
    \subfloat[The ``ghost'' circle starts approaching]{%
        \includegraphics[width=0.24\textwidth]{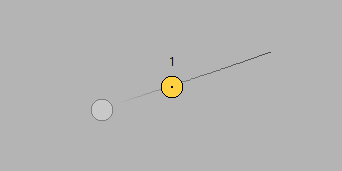}
    }
    \subfloat[The ``ghost'' circle disappears]{%
        \includegraphics[width=0.24\textwidth]{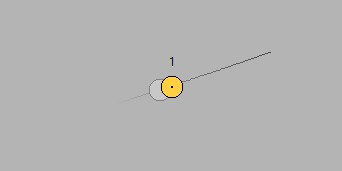}
    }
    \subfloat[A line indicates the movement direction]{%
        \includegraphics[width=0.24\textwidth]{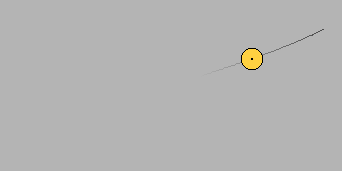}
    }

    \vspace{0.2cm}

    Saccades Movement
    \subfloat[Countdown at the start]{%
        \includegraphics[width=0.24\textwidth]{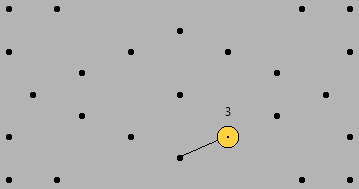}
    }
    \subfloat[The target shrinks]{%
        \includegraphics[width=0.24\textwidth]{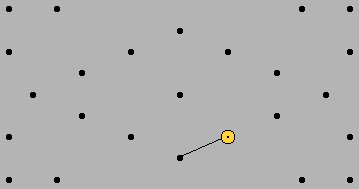}
    }
    \subfloat[The target shrinks again]{%
        \includegraphics[width=0.24\textwidth]{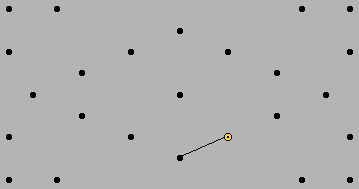}
    }
    \subfloat[The target jumps]{%
        \includegraphics[width=0.24\textwidth]{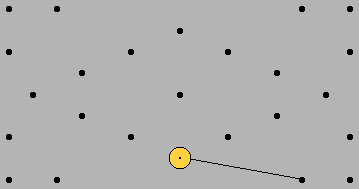}
    }
    \caption{Stimuli presentation. Top: Presentation of the stimulus in the level-2 smooth experiment. Bottom: Presentation of the stimulus in the level-2 saccades experiment.}
    \label{fig:dataset:stimuli}
\end{figure}

Several design choices were made to help the participants anticipate the target's movement, to minimize the distance between the position of the target and the participant's gaze.
The presentation differed between the ``smooth'' and ``saccades'' paradigms. In the ``smooth'' experiments, in addition to the yellow circle moving across the screen as the target, a line was displayed to show the expected direction of the target's movement. Additionally, before the target started moving, a ``ghost'' circle appeared, indicating the start and speed of the movement. See Figure~\ref{fig:dataset:stimuli} for an example of the stimuli presentation in a ``smooth'' experiment.

In the ``saccades'' experiments, a line connected the current position of the target with its next position, i.e., the position to which the target jumped. 
To indicate the moment when the jump occurred, the circle got scaled down $3$ times, completely disappearing the last time. The moment the circle disappeared was the moment the jump happened. This is shown in the lower part of Figure~\ref{fig:dataset:stimuli}.

\subsection{Data Preprocessing}
\label{sec:dataset:preprocessing}

The suggested preprocessing pipeline consists of two steps, (i) imputation of missing values, and (ii) band-pass filtering. While the first step is computationally expensive, the second is efficient. To facilitate use of the data set, pre-processed data with imputed missing values is provided alongside the raw data and code to filter it.

\subsection*{Missing Value Imputation}
Missing values were recorded as zeros, which could also occur as valid values. Thus, the first step was to distinguish actual missing values from legitimate zero readings. To address this, zero values that occurred as part of an uninterrupted sequence of at least three zeros were identified as missing values. That means, a zero was flagged as missing if it was preceded, followed, or surrounded by two other zeros. This approach was adopted to ensure that isolated zero values were retained as valid data points, while sequences of zeros (which are unlikely to occur naturally in the data) were treated as missing data.

\begin{figure}
    \centering
    \includegraphics[width=0.8\textwidth]{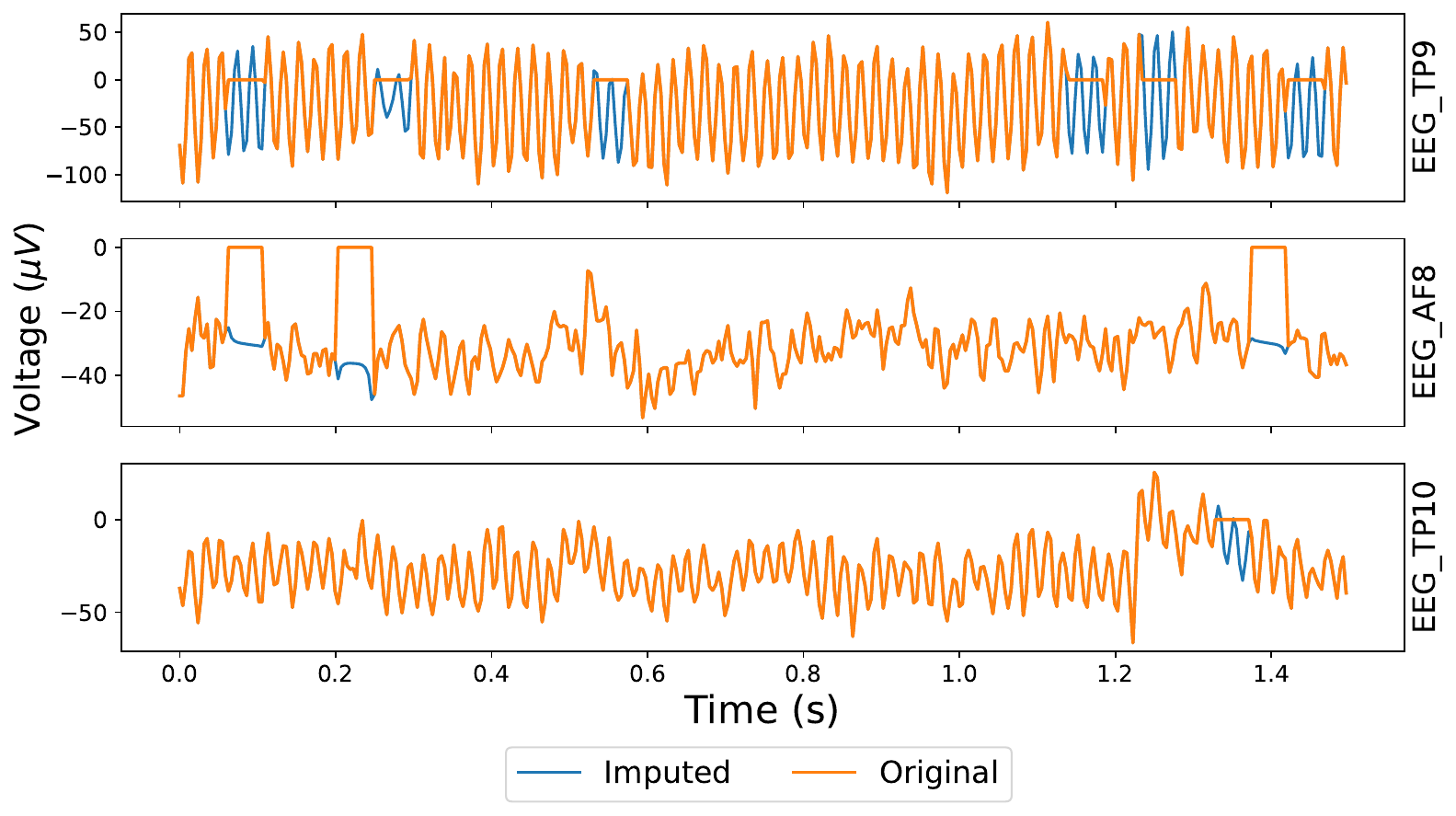}
    \caption{Exemplary imputation of missing values for recording ``P$045$\_$01$ level-$1$-saccades''. Electrode AF7 did not contain missing values in this segment and is not displayed.}
    \label{fig:methods:missing-value-imputation}
\end{figure}

Once the missing values were identified, they were imputed using a Kalman smoother. In this case, a Seasonal Autoregressive Integrated Moving Average
(SARIMA) model was used as the underlying state-space model for the Kalman smoother. The optimal SARIMA model for each recording was determined automatically using the \verb|auto_arima| function from the \verb|pmdarima| library.
This function performs differencing tests to decide the order of differencing required for stationarity, together with a stepwise algorithm, as outlined by \cite{hyndman2008}, to select the $p$, $q$, and seasonal $P$ and $Q$ orders based on the Akaike Information Criterion. 
The \verb|auto_arima| function was applied with default parameters, with the exception of the seasonal lag, which was set to $5$. 
This choice was made due to the particularly strong presence of the \SI{50}{Hz} power line noise signal in the data, which, at a sampling rate of \SI{256}{Hz}, corresponds to a period of approximately $5$ samples ($256 / 50 = 5.12 \approx 5$). Figure~\ref{fig:methods:missing-value-imputation} illustrates the effect of the missing value imputation on the data.

\subsection*{Filtering}
As EEG data is highly prone to noise from various sources, filtering is a common preprocessing step in its analysis. Multiple frequency filters were applied subsequently. First, a \SI{60}{Hz} notch filter was applied to remove the noise from the monitor refresh rate. This was followed by another \SI{50}{Hz} notch filter to remove the noise from the power lines. Finally, a \SI{0.5}{Hz} to \SI{40}{Hz} Butterworth bandpass filter was applied to remove noise from other sources, such as muscle activity and, baseline drift. All filters were applied forwards and backwards to avoid phase distortion.

\begin{figure}
    \centering
    \includegraphics[width=0.8\textwidth]{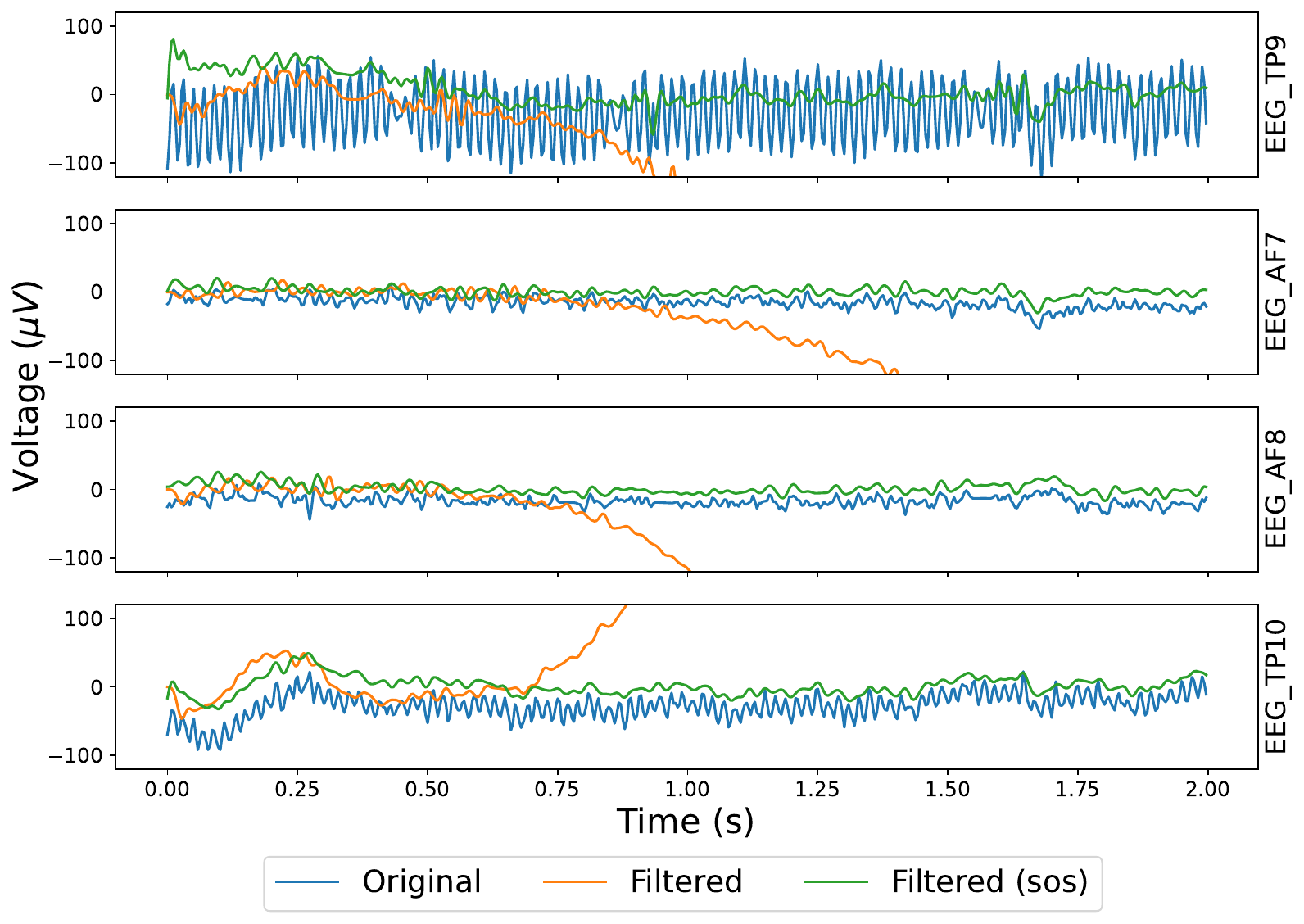}
    \caption{Frequency filters applied to the recording ``P$045$\_$01$ level-$2$-saccades'', where missing values were imputed as described in the above section.}
    \label{fig:methods:filtering:results}
\end{figure}

The filters were implemented using second-order sections (SOS) instead of directly applying the difference equation, as this was found to be more stable during testing. In Figure~\ref{fig:methods:filtering:results} the effect of applying the filters on the data is shown. It is evident that the vanilla filter causes the signal to diverge over time, while the SOS-filter keeps the signal stable. Both filters used the combination of \SI{60}{Hz} and \SI{50}{Hz} notch and \SI{0.5}{Hz} to \SI{40}{Hz} bandpass filter. However, the vanilla filter used an $8$th order bandpass filter and applied all filters only forwards, as applying them backwards would cause the signal to attain extreme values throughout. The SOS-filter used a $4$th order bandpass filter and applied all filters both forwards and backwards, doubling the order of the filter and correcting phase shift.

\begin{figure}
    \centering
    \includegraphics[width=0.8\textwidth]{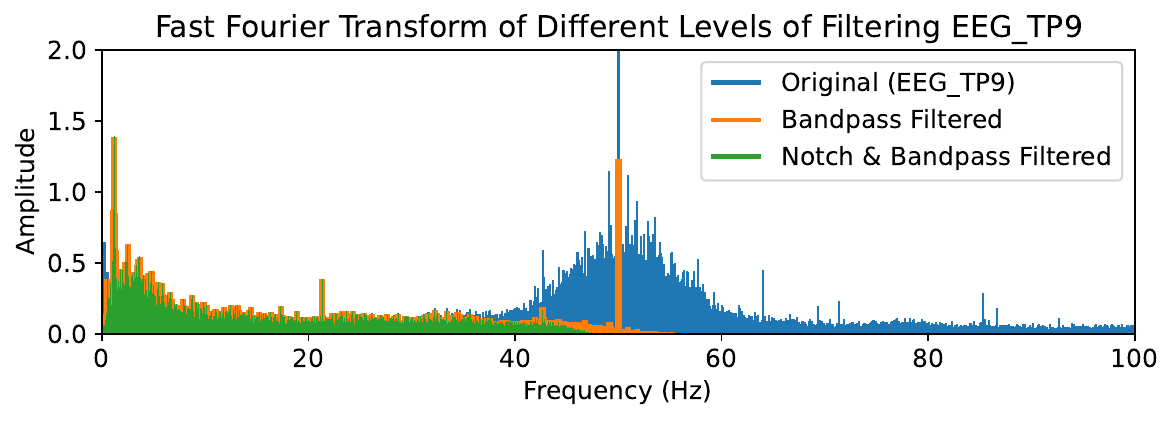}
    \caption{Bandpass filter applied to electrode TP$9$ of recording "P$045$\_$01$ level-$2$-smooth".}
    \label{fig:methods:filtering:bandpass-insufficient}
\end{figure}

Even though the bandpass filter limits the data to the frequency range from \SI{0.5}{Hz} to \SI{40}{Hz} --- which should exclude the \SI{50}{Hz} and \SI{60}{Hz} noise --- we have found that applying the bandpass alone was insufficient to remove the noise entirely. This can be seen in Figure~\ref{fig:methods:filtering:bandpass-insufficient} where the noise from the power lines is still present in the data after applying the bandpass filter.

\section{Data Records} 

In total, the dataset contains $11$ hours and $45$ minutes of EEG and eye tracking data. The total time recorded per paradigm and the duration in one session can be seen in Table~\ref{tab:dataset:duration}.

The dataset is available as a single denormalized \texttt{CSV} file, which contains all the data collected during the sessions. The \texttt{CSV} file has a size of \SI{1.42}{GB} and the following columns: \verb|Participant_no|, \verb|Task|, \verb|Session_no|, \verb|Timestamp|, \verb|EEG_TP9|, \verb|EEG_AF7|, \verb|EEG_AF8|, \verb|EEG_TP10|, \verb|Gaze_x|, \verb|Gaze_y|, \verb|Stimulus_x|, \verb|Stimulus_y|. Any samples recorded before or after the stimulus presentation are removed.

In addition to the \texttt{CSV} file, a ``csv'' and an ``xdf'' folder are provided. Inside both folders, there are four subfolders, one for every experimental paradigm. Furthermore, every subfolder contains two more folders called ``train'' and ``test'', which contain one file per recording. When machine learning models are created to predict the true eye position from the EEG data, only the training data should be used for fitting the model, while the test data should be used to evaluate the fitted model. The filename always follows the naming scheme PXXX\_YY.csv or PXXX\_YY.xdf, where XXX is the participant number and YY is the session number for this participant. 
Writing out the folder structure in a tree-like manner, one would get the following:

\vspace{5pt}
\parbox{\textwidth}{
	\dirtree{%
		.1 (csv or xdf).
		.2 level-1-saccades.
		.3 train.
		.4 P001\_01.(csv or xdf).
		.4 P001\_02.(csv or xdf).
		.4 P002\_01.(csv or xdf).
		.4 \ldots.
		.3 test.
		.4 P037\_01.(csv or xdf).
		.4 P042\_01.(csv or xdf).
		.4 \ldots.
		.2 level-1-smooth.
		.3 \ldots.
		.2 level-2-saccades.
		.3 \ldots.
		.2 level-2-smooth.
		.3 \ldots.
	}
}
\vspace{5pt}

In the ``csv'' folder, the files are \texttt{CSV} files with columns 
\verb|Timestamp|, \verb|EEG_TP9|, \verb|EEG_AF7|, \verb|EEG_AF8|, \verb|EEG_TP10|, \verb|Gaze_x|, \verb|Gaze_y|, \verb|Stimulus_x|, \verb|Stimulus_y|. A detailed description of the columns is provided in Table \ref{tab:dataset:description}.

The ``xdf'' folder contains \texttt{XDF} files with the raw recorded data. Those files contain additional metadata of the multiple data streams. Users interested in more detailed, time-series structured data may find the \texttt{XDF} files useful. 

The stimuli presentation software sends the current program status via a LSL stream. In particular, this includes a start and end marker. The first line of the \texttt{CSV} file is the first measurement after the start marker. Each line corresponds to a further EEG value. Since other signals (gaze position, program state, position of the target) do not change with the same frequency as the EEG data is recorded, these values always correspond to the most recent ones at this point in time. The last line of the \texttt{CSV} file is the last measurement before the end marker. In addition, the absolute positions in px from the \texttt{XDF} file are converted to positions relative to the center of the bounding box in mm.

Both the \texttt{CSV} file containing the complete dataset, and the ``csv'' and ``xdf'' folders are freely available on Zenodo: \url{https://zenodo.org/records/14860668}.

\begin{table}
    \centering
        \begin{tabular}{l|l|l|l}
            \toprule
            Source & Column Name & Data Type & Description \\
            \midrule
            LSL & Timestamp & float & Timestamp from synchronized data streams \\
            \midrule 
            \multirow{4}{*}{EEG} & EEG\_TP9 & float &  EEG signal from electrode TP9 \\
            & EEG\_AF7 & float &  EEG signal from electrode AF7 \\
            & EEG\_AF8 & float &  EEG signal from electrode AF8 \\
            & EEG\_TP10 & float &  EEG signal from electrode TP10 \\
            \midrule 
            \multirow{2}{*}{Eye Tracker} & Gaze\_x & float & (estimated) $x$ coordinate of the gaze on screen \\ 
            & Gaze\_y & float & (estimated) $y$ coordinate of the gaze on screen \\ 
            \midrule 
            \multirow{2}{*}{Presentation App} & Stimulus\_x & float & $x$ coordinate of the target on screen \\
            & Stimulus\_y & float & $y$ coordinate of the target on screen \\
            \bottomrule
        \end{tabular}
    \vspace{5pt}
    \caption{Data structure description.}
    \label{tab:dataset:description}
\end{table}

\section{Technical Validation} 

In this section, we describe the technical validation of the data collected using a consumer-grade BCI headset and a webcam-based eye tracker. Given the nature of the experiments, there was a consistent trade-off between obtaining high-quality data and maintaining a realistic, non-laboratory setting. While the consumer-level hardware offered the advantage of a more natural and accessible data collection environment, it also posed challenges related to signal accuracy and reliability. This section outlines the steps taken to assess the quality of the data and address potential issues inherent in using such devices outside of controlled lab conditions.

\subsection{Known Quality Issues} \label{sec:technical-validation:quality-issues}

During the data recording process, live EEG measurements and gaze data (tracked via the webcam) were continuously monitored. This real-time monitoring enabled early detection of recording issues, allowing the task to be restarted when anomalies were observed. In certain cases, it was not possible to achieve reliable readings from specific electrodes. These instances, and similar issues, were documented during the recording sessions, and the corresponding recordings might be excluded from subsequent experiments. 
The full list of $14$ recordings, with known quality issues, is shown in Table~\ref{tab:technical-validation:manual-exclusion}. 
\begin{table}
    \centering
    \begin{tabular}{l l}
        \toprule
        Recording(s) & Reason for Exclusion \\
        \midrule
        P$002$\_$01$ & Webcam calibration quality declined significantly \\
        P$004$\_$01$ & \texttt{Gaze\_x} attained unrealistically high values \\
        P$016$\_$01$ -- P$020$\_$01$ & A lot of missing values due to Bluetooth interference \\
        P$050$\_$01$ level-$1$-saccades & EEG disconnected before the recording finished \\
        P$062$\_$01$ -- P$067$\_$01$ & Very high AF$8$ readings due to faulty hardware \\
        P$079$\_$01$ & High TP$9$ and TP$10$ readings from wearing a headgear \\
        \bottomrule
    \end{tabular}
    \vspace{5pt}
    \caption{List of recordings with known quality issues..}
    \label{tab:technical-validation:manual-exclusion}
\end{table}

\subsection{EEG Data}

The aim of the experiments was to create a representative dataset of EEG data recorded with consumer-grade hardware in a realistic environment. For the EEG data, we had to carefully decide when the signal-to-noise ratio became too low, even for this purpose. At the beginning of each recording, the signal quality was assessed visually. The EEG data was visualized using a viewer from the Python library \texttt{muselsl} \citep{muselsl}, which applies a bandpass filter between 1 and 40 Hz by default. The filtered signal was then checked to ensure it was primarily below \SI{30}{\micro \volt}. This naïve heuristic helped identify recordings whose signal-to-noise ratio was too low even for a consumer-level BCI headset. A more systematic approach was avoided in order to collect representative and realistic EEG signals.

The data contains missing values due to incorrect transmission, which were stored as zeros. In Section \ref{sec:dataset:preprocessing}, we explain the approach that was used for the imputation of missing values.

\subsection{Stimulus Presentation}

Several measures were implemented, to ensure that participants could effectively follow the target on the screen during the experiment. As described in Section \ref{sec:dataset:presentation}, the direction and timing of the target's movement could be easily predicted due to visual cues. The target was displayed as a circle that got scaled down three times before jumping to the next position. This gradual reduction in size should help the participants to better estimate the time of the next jump. 
Further visual cues allowed participants to anticipate the next location or direction of movement, such as a ``ghost'' dot and a line hinting in the latter direction, cf. Figure \ref{fig:dataset:stimuli}. 

In addition to the visual cues, the participants tried out the various tasks in a trial run before the recording. This ensured that participants were familiar with the target's behavior and could follow it better.

These measures were crucial for maintaining the integrity of the data, while finding a balance between realistic conditions and the need for accurate eye tracking. By making the target visually clear, predictable, and easy to follow, we aimed to optimize the participant's ability to complete the task successfully, thus improving the reliability of the recorded data.

\subsection{Eye Tracking Data}

The gaze position, as obtained from the eye tracking software, should not be mistaken for the ``ground truth'' of eye movements. The data was captured using webcam-based eye tracking software, which, although effective for tracking gaze in non-laboratory settings, inherently introduces certain inaccuracies and limitations.

Generally, the accuracy of webcam-based eye tracking is influenced by factors such as lighting conditions, camera angle, and the calibration process. These factors can lead to discrepancies between the participant’s actual gaze position and the data recorded by the software. The accuracy of the used eye-tracking software was reported to be between 1.4° and 1.9° \citep{heck2023, falch2024}. When considering a distance between participant and monitor of approximately 60\,cm, this translates to inaccuracies of 15\,mm to 20\,mm. This inaccuracy might be amplified by external factors, such as head movement or varying lighting. 

Moreover, a lag between the participant’s real-time eye movements and the tracked data can be introduced by the software and hardware used. While this lag is generally small, it can introduce substantial differences, particularly when tracking fast movements. 

These limitations are inherent to the experiment based on consumer-grade software and hardware, and the eye tracking data must be interpreted accordingly.

\subsection{Correlation Between Target and Gaze} \label{sec:technical-validation:cross-correlation}

\begin{figure}
    \centering
    \includegraphics[width=0.9\textwidth]{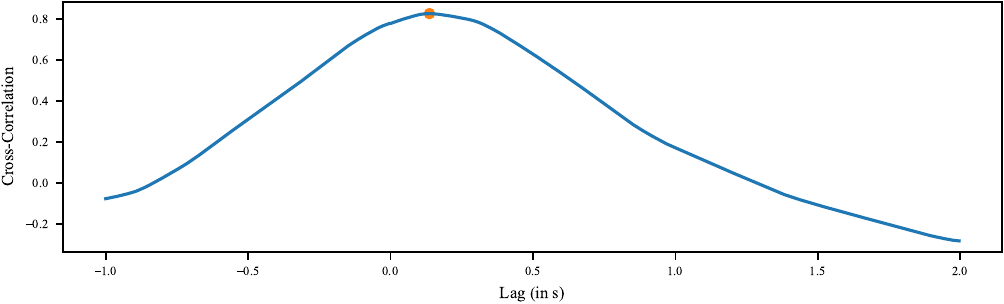}
    \caption{Cross-correlation function between $y$ coordinates of gaze and target for the level-1 ``saccades'' recording of participant 21.}
    \label{fig:validation:ccf}
\end{figure}

The target positions correspond exactly to the location of the target on the screen, while the gaze positions are estimated by the eye tracking software. It is anticipated that the gaze will lag behind the target. To validate this relationship, we calculated the cross-correlation functions between the $x$ and $y$ coordinates of the gaze and target, respectively (see, e.\,g., Figure \ref{fig:validation:ccf}). Subsequently, we calculated the lag of maximal cross-correlation for each recording. The averages of these lags across all recordings are reported in Table \ref{tab:validation:lag}.

\begin{table}
    \centering
    \begin{tabular}{l|l|l}
        \toprule
        Experimental Paradigm   & Avg. lag ($x$) & Avg. lag ($y$) \\
        \midrule
        level-1-saccades & \SI{0.18}{\second} ($\pm$ \SI{0.13}{\second}) & \SI{0.23}{\second} ($\pm$ \SI{0.13}{\second}) \\
        level-1-smooth   & \SI{0.36}{\second} ($\pm$ \SI{0.20}{\second}) & \SI{0.38}{\second} ($\pm$ \SI{0.20}{\second}) \\
        level-2-saccades & \SI{0.16}{\second} ($\pm$ \SI{0.20}{\second}) & \SI{0.19}{\second} ($\pm$ \SI{0.18}{\second}) \\
        level-2-smooth   & \SI{0.26}{\second} ($\pm$ \SI{0.16}{\second}) & \SI{0.32}{\second} ($\pm$ \SI{0.18}{\second}) \\
        \bottomrule
    \end{tabular}
    \vspace{5pt}
    \caption{Average lag between target and gaze movement, as determined via their cross-correlation function, across all recordings. The respective standard deviation across all recordings is provided in parentheses.}
    \label{tab:validation:lag}
\end{table}

\section{Usage Notes} 

When working with the dataset, there are several aspects to take into account. EEG data is highly noisy even when recorded in a laboratory. The signals measured with consumer-grade hardware have a low signal-to-noise ratio. Therefore, it is recommended to preprocess the data first, or use the provided preprocessed data. Exemplary code to load the preprocessed data can be found in the GitHub repository: \url{https://github.com/FlorianHeinrichs/eeg_eye_tracking}.

To get started with the dataset, we recommend using \texttt{CSV} files. For users who are  familiar with the \texttt{XDF} format, we recommend using the latter, as it contains additional information and metadata. 

When the dataset is used to train and evaluate machine learning models, only the training data should be used for model training, model selection and hyperparameter tuning. The test data should only be used for a final evaluation, and, in particular, not for model selection or hyperparameter tuning. For the latter tasks, the training data might be split into ``training'' and ``validation'' data.

We assume that methods from time series analysis are suitable for interpreting the data. With regard to the ``smooth'' experiments, we expect that methods of functional data analysis, such as functional neural networks \citep{heinrichs2023} or transform-invariant functional PCA \citep{heinrichs2024}, could be particularly helpful, as they take the smoothness of the data into account.

\section{Code Availability}

The code used in this study is written in Python and is publicly available on GitHub (\url{https://github.com/FlorianHeinrichs/eeg_eye_tracking}). The code was developed with Python (3.11), Matplotlib (3.9.2), NumPy (1.26.4), Pandas (2.2.3), Pooch (1.8.2), SciPy (1.14.1), Statsmodels (0.14.4), and respective dependencies.

The repository contains the following key components:
\begin{itemize}
    \item \texttt{stimuli-presentation-app}: Code for the application used to record the data.
    \item \texttt{analyse\_data.py}: Script to calculate the lag between gaze and stimulus through the cross-correlation function, as described in Section \ref{sec:technical-validation:cross-correlation}.
    \item \texttt{impute\_missing\_values.py}: Script to impute missing values, as described in Section \ref{sec:dataset:preprocessing}
    \item \texttt{xdf\_to\_csv.py}: Script to convert the raw data from XDF to CSV format.
\end{itemize}

Most importantly, the repository includes the \texttt{load\_data} module, which provides functions to load and filter the data. The \texttt{load\_data} functions leverage the \texttt{pooch} library in the backend to automatically fetch the required dataset from Zenodo, when it is called for the first time.

\section{Author Contributions} 
T.A., F.H. conceived the experiments. T.A. developed the software and conducted the experiments. F.H. contributed the modules \texttt{load\_data} and \texttt{analyze\_data}. T.A., F.H. performed statistical analysis. All authors contributed to the writing of this manuscript, critically reviewed it and approved the final version.

\section{Competing Interests} 

The authors declare no competing interests.

\bibliographystyle{apalike}
\bibliography{bibliography}

\end{document}